\documentstyle[bal,times,psfig]{article}
\ifx \htmlstyloaded\relax  \else\let\htmlstyloaded\relax\fi

\newcommand{\htmladdnormallink}[2]{#1}

\newcommand{\htmladdimg}[1]{}

\newcommand{\externallabels}[2]{}

\newcommand{\externalref}[1]{}

\makeatletter
\def\makeinnocent#1{\catcode`#1=12 }
\def\csarg#1#2{\expandafter#1\csname#2\endcsname}

\def\ThrowAwayComment#1{\begingroup
    \def\CurrentComment{#1}%
    \let\do\makeinnocent \dospecials
    \makeinnocent\^^L
    \endlinechar`\^^M \catcode`\^^M=12 \xComment}
{\catcode`\^^M=12 \endlinechar=-1 %
 \gdef\xComment#1^^M{\def\test{#1}
      \csarg\ifx{PlainEnd\CurrentComment Test}\test
          \let\html@next\endgroup
      \else \csarg\ifx{LaLaEnd\CurrentComment Test}\test
            \edef\html@next{\endgroup\noexpand\end{\CurrentComment}}
      \else \let\html@next\xComment
      \fi \fi \html@next}
}
\makeatother

\def\includecomment
 #1{\expandafter\def\csname#1\endcsname{}%
    \expandafter\def\csname end#1\endcsname{}}
\def\excludecomment
 #1{\expandafter\def\csname#1\endcsname{\ThrowAwayComment{#1}}%
    {\escapechar=-1\relax
     \csarg\xdef{PlainEnd#1Test}{\string\\end#1}%
     \csarg\xdef{LaLaEnd#1Test}{\string\\end\string\{#1\string\}}%
    }}

\excludecomment{comment}

\excludecomment{rawhtml}

\excludecomment{htmlonly}
\newcommand{\html}[1]{}

\newcommand{\hyperref}[4]{#2\ref{#4}#3}

\newcommand{\htmlimage}[1]{}

\newcommand{\htmladdtonavigation}[1]{}

\begin{document}
%
%
\title{Network Management of Predictive Mobile Networks\thanks{This
paper is partially funded by ARPA contract number J-FBI-94-223 and
Sprint under contract CK5007715.}}
\author{\htmladdnormallink{Stephen F. Bush}
{http://www.tisl.ukans.edu/\~{}sbush}, Victor S. Frost
\from
Information and Telecommunication Technologies Center \\
Department of Electrical Engineering \& Computer Science\\
University of Kansas\\
Lawrence, KS 66045-2228 \\
\email{sbush@tisl.ukans.edu}
}
\markright{S. Bush, V. Frost
/ Network Management of Predictive Mobile Networks}
\maketitle
\begin{abstract}

There is a trend toward the use of predictive systems in communications
networks. At the systems and network management level predictive
capabilities are focused on anticipating network faults and performance
degradation. Simultaneously, mobile communication networks are being
developed with predictive location and tracking mechanisms. The
interactions and synergies between these systems present a new set of
problems. A new predictive network management framework is developed and
examined.  The interaction between a predictive mobile network and the
proposed network management system is discussed.  The Rapidly Deployable
Radio Network is used as a specific example to illustrate these
interactions.\footnote{This paper is partially funded by ARPA
contract number J-FBI-94-223 and Sprint under contract CK5007715.}

{\bf Keywords: {\em Prediction Mobile Network Management Time Warp
Virtual Network Configuration}}

\end{abstract}

%
%
\section{INTRODUCTION}

Recently proposed mobile networking architectures and protocols
involve predictive mobility management schemes.  For example,	
an optimization to a Mobile IP-like protocol using IP-Multicast is
described in \cite{Seshan}. Hand-offs are anticipated and data is
multicast to nodes within the neighborhood of the predicted handoff.
These nodes intelligently buffer the data so that no matter where
the mobile host (MH) re-associates after a handoff no data will be lost.
Another example \cite{Liu, Liuphd} proposes deploying mobile floating 
agents that decouple 
services and resources from the underlying network. These agents are
pre-assigned and pre-connected to predicted user locations.
This paper focuses on the Rapidly Deployable Radio
Networks Project \cite{BushICC96, BushRDRN} as an example
of a specific predictive mobile network. The Virtual Network Configuration
algorithm developed as part of RDRN uses a predictive mechanism 
for every phase of configuration, including location and handoff.

Progress is being made in research involving predictive system
and network management \cite{Schiaffino}. This paper develops 
a variation of the Virtual Network Configuration Algorithm as proposed 
for the RDRN \cite{BushRDRN} for a predictive network management system. 
The predictive capability of such a system can be used to help optimize 
its own operation by controlling the management of the polling rate. 
Finally, a discussion
of how predictive mobile networks and predictive network management
systems should interact is presented.

\section{A PREDICTIVE NETWORK MANAGEMENT SYSTEM}

Systems management means the management of 
heterogeneous subsystems of network devices, processing platforms,
distributed applications, and other components found in
communications and computing environments. 
Current system management relies on presenting a model to the user of the
managed system that should accurately reflect the current
state of the system and should ideally be capable of predicting the 
future health of the system.
System management relies on a combination of asynchronously generated
alerts and polling to determine the health of a system
\cite{Steinber91}.

The management application presents state information such as link state, 
buffer fill, and packet loss to the user in the form of a model \cite{Sylor}.
The model can be as simple as a passive display of nodes on a screen, or
a more active model that allows displayed nodes to change color based 
on state changes, 
or react to user input by allowing the user to manipulate the nodes,
causing values to be set on the managed entity.
This model can be made even more active by enhancing it with
predictive capability.
This enables the management system to manage itself, for example,
to optimize its polling rate. The two major management protocols,
Simple Network Management Protocol (SNMP) \cite{SNMP} and Common 
Management Information Protocol (CMIP) \cite{CMIP}, allow the management 
station to poll a managed entity to determine its state.
To accomplish real-time and predictive network management
in an efficient manner, the model should
be updated with real-time state information when it becomes available,
while other parts of the model work ahead in time. Those objects
working ahead of real-time can predict
future operation so that system management parameters such as polling times 
and thresholds can be dynamically adjusted and problems can be anticipated. 
The model will not deviate too far from reality because those processes
found to deviate beyond a certain threshold are rolled
back, as explained in detail later.
The processes' messages must obey the following rules for consistency
in \cite{Lamport78}:
\newtheorem{lamportr}{Rule}
\begin{lamportr}
If two events are scheduled for the same process, then the event with
the smaller timestamp must be executed before the one with the larger
timestamp.
\end{lamportr}
\begin{lamportr}
If an event executed at a process results in the scheduling of another
event at a different process, then the former must be executed before the
latter.
\end{lamportr}

To determine the characteristics and performance of
this predictive network management algorithm, we will review the research on
performance and modeling of other lookahead algorithms and Time Warp
in particular.
A comparison of the conservative Chandy-Misra 
approach and the optimistic Time Warp
is presented in \cite{Felderman90}. This is done using a
typical queuing theory approach which assumes exponential service times.
There have been several other detailed comparisons between conservative and 
optimistic 
methods of simulation. These studies also make simplifying assumptions. 
In \cite{Lin90a},
it is shown that in a feed-forward network, the time of execution of a message
will occur earlier in Virtual Time than its corresponding message in 
the synchronous parallel algorithm described in \cite{Lamport78}.
In \cite{Lipton90}, it is shown that Time Warp can out-perform the 
conservative technique known as Chandy-Misra by a factor of
$P$, $P$ being the number of processors, but that no such
model\index{simulation!model} in which Chandy-Misra\index{Chandy-Misra}
out-performs Time Warp\index{Time Warp} by a factor the number of
processors used exists.
Past work has examined the performance of Time Warp by comparing it to 
conservative mechanisms
\cite{Lin90a}
or simulating the Time Warp mechanism itself \cite{Turnbull92}.
In this paper the focus is not only on analyzing and optimizing 
speed of execution but also using the algorithm to maintain 
network management prediction accuracy.

One goal of this research is to minimize polling overhead
in the management of large systems \cite{Takagi86}.
Instead of basing the polling rate on the characteristics of the
data itself, the entity is emulated some time into the future 
to determine the characteristics of the data to be polled.
Polling is still required with this predictive network management
system to verify the accuracy of the emulation. 

\section{RELATIONSHIP BETWEEN NETWORK MANAGEMENT AND PARALLEL DISCRETE
EVENT SIMULATION}

Management information from standards-based managed
entities must be mapped into this predictive network management system.
Network management systems rely upon standard mechanisms to obtain the state of 
their managed entities in near real-time. These mechanisms, 
SNMP \cite{SNMP} and CMIP \cite{CMIP} for example, use both solicited
and unsolicited methods. The unsolicited method uses
messages sent from a managed entity to the manager. These unsolicited
messages are
called traps or notifications; the former are not acknowledged while
the latter are acknowledged. These messages are very similar to messages
used in distributed simulation algorithms; they contain a timestamp and 
a value, they are sent to a particular destination, i.e. a management
entity, and they are the result of an event which has occurred.

Information requested by the management system from a particular managed
entity is solicited information. It also corresponds to messages 
in a distributed simulation. It provides a time and a value; however,
not all such messages are equivalent to messages in distributed
simulation. These messages 
provide the management station with the 
current state of the managed entity, even though no change of 
state may have occurred or multiple state changes may have occurred. 
The design of a management system that requests information concerning the
state of its managed entities at the optimum time has always been a problem
in network management. If management information is requested too frequently, 
bandwidth is wasted, if not requested frequently enough, critical state 
change information will be missed.

We will assume for simplicity that each managed entity is represented in
the predictive management system by a Logical Process. It would greatly 
facilitate system management if vendors provide not only the standards based 
SNMP MIBs as they do now, but also standard 
simulation code that models entity or application behavior and can be 
plugged into the management system just as Management Information Bases are used today. Vendors 
have such simulation models of their managed devices readily available from 
product development
\section{THE PREDICTIVE NETWORK MANAGEMENT SYSTEM  ALGORITHM}

Terminology borrowed from previous distributed simulation algorithms
has a slightly different meaning in this predictive network 
management system\index{predictive network management system}. In addition,
new terminology must be introduced. Thus it is important that the terminology 
be precisely defined. 

The predictive network management system management system algorithm 
encapsulates Physical Process
simulating managed network devices within an Logical Process.
A Physical Process is nothing more than the executing process defined by the 
program code.
An Logical Process consists of the Physical Process and additional data structures 
and instructions to maintain message 
order and correct operation as the system executes ahead of real 
time. An Logical Process contains a Receive Queue, Send Queue, and
State Queue. The Receive Queue maintains newly arriving messages in order by 
their Receive Time. 
The Send Queue maintains copies of previously
sent messages in order of their send times. The state of the Logical Process is
periodically saved in the State Queue. The Logical Process also contains its notion of
time known as Local Virtual Time and a Tolerance 
($\Theta$)
that is the allowable deviation between actual and predicted
values of incoming messages. Also, the Current State
of a Logical Process will be the current state of the Logical Process and its 
encapsulated Physical Process.
The predictive network management 
system\index{predictive network management system} 
contains a notion of the complete system time 
known as Global Virtual Time and a 
sliding window known as the {\bf Lookahead\index{Lookahead}} time 
($\Lambda$\index{$\Lambda$|see{Look Ahead}}).
Messages contain the Send Time, Receive Time, Anti-toggle\index{Anti-toggle},
and the message contents. The Receive Time is the time this message should be received by
the destination Logical Process\index{Logical Process}. The Send Time is the time this message was sent by the
originating Logical Process\index{Logical Process}. The Anti-toggle field is the anti-toggle and is used for 
creating an anti-message\index{anti-message} to remove the effect of 
false messages\index{false message} as described
later. A message will also contain a field for the current 
Real Time. This is used to differentiate a real 
message\index{real message} from a virtual message\index{virtual message}.

A {\bf driving process\index{predictive network management system!driving process}} is required to predict 
future events and inject them into the system. For example, in a mobile
system such as the Rapidly Deployable Radio Network \cite{BushRDRN}, 
the Global Positioning System is used to provide each node with
its current position.
The Global Positioning System receiver process runs in real-time\index{real-time}
and inject future predicted location messages. 
In the predictive network management system, the driving process may be
the number of expected users and their estimated bandwidth usage.
The driving process(es)\index{predictive network management 
system!driving process} originate virtual messages via internal prediction.
The remaining Physical Processes\index{PP} react to these messages 
as though they are real messages\index{real message}.
A message which is generated and time-stamped with the current time will
be called a {\bf real message\index{real message}}. Messages which 
contain future event information and are time-stamped with a time greater 
than current time are called {\bf virtual messages\index{virtual message}}. 
If a message arrives at a Logical Process\index{LP} out-of-order or with invalid 
information, it is called a {\bf false message\index{false message}}. 
A false message\index{false message} causes an 
Logical Process\index{LP} to rollback\index{rollback}.

Rollback\index{predictive network management system!rollback} is a 
mechanism by which a Logical Process\index{LP} returns to a known correct state.
The rollback\index{rollback} occurs in three phases. In the first phase, 
the Logical Process\index{LP} state
is restored to a time strictly earlier than the time stamp of the
false message\index{false message}. In the second phase, anti-messages are 
sent to cancel the effects of invalid messages that had been generated 
before the arrival of the false message\index{false message}. An {\bf 
anti-message\index{anti-message}} contains exactly the
same contents as the original message with the exception of an anti-toggle
bit\index{Anti-toggle!anti-toggle bit} that is now set. When the 
anti-message\index{anti-message} and original message meet,
they are both annihilated\index{annihilation}. The final phase consists 
of executing the Logical Process\index{Logical Process} forward in time from 
its rollback\index{rollback} state to the time the 
false message\index{false message} arrived. No messages are canceled or sent 
between the time to which the Logical Process\index{Logical Process} rolled 
back and the time of the false message\index{false message}. 
Because these messages are correct there is
no need to cancel or re-send them. This increases performance, and it 
reduces the number of messages causing roll-back. Note that another 
false message\index{false message} or anti-message\index{anti-message} 
may arrive before this final phase has completed without causing any problems.
\section{CHARACTERISTICS OF THE PREDICTIVE NETWORK MANAGEMENT SYSTEM}

There are two types of false messages generated in this predictive 
network management system; those produced by messages arriving in the past 
Local Virtual Time of an Logical Process and those produced because the Logical Process is
generating results which do not match reality. If rollbacks occur for
both reasons the question arises as to whether system will be
stable. A stable predictive network management system is one in which
rollbacks do not have a significant impact on system performance.
A stable system is able to make reasonably accurate predictions far 
enough into the future to be useful.
An unstable system will have its performance degraded by rollbacks
to the point where it is not able to predict ahead of real-time.
Initial results, shown later, indicate that predictive 
network management systems using the algorithm described in this paper
can be stable.

There are several parameters in this predictive network management system 
which must be determined. The first is how often the predictive network 
management system should check the Logical Process to verify that past results match 
reality. There are two conditions which cause Logical Processes in the system to 
have states which differ from the system being managed and to
produce inaccurate predictions. The first is that the predictive
model which comprises an Logical Process is most likely a simplification of an
actual managed entity and thus cannot model the entity with perfect
fidelity. The second condition occurs when events outside the scope of the
model may occur which lead to inaccurate results. However, a benefit
of this system is that it self-adjusts to both of these conditions.

The optimum choice of verification 
query time, $T_{query}$, is important because querying entities is
something the predictive management system should minimize while still 
guaranteeing that the
accuracy is maintained within some predefined tolerance, $\Theta$. 
For example, the network management station may predict user location
as explained later. If the physical layer attempts spatial 
reuse via antenna beamforming
techniques as in the Rapidly Deployable Radio Network project, then there is an acceptable amount
of error in the steering angle for the beam and thus the node location
is allowed to be within a tolerance.
The tolerances are set for each state variable 
sent from a Logical Process. State verification can be done in one of at least
two ways. The Logical Process state can be compared with previously saved states
as real time catches up to the saved state times or output message
values can be compared with previously saved output messages in the
send queue.
In the prototype implemented for this predictive network management 
system state verification is done based on states saved in the state
queue.  This implies that all Logical Process states must be saved from the Logical Process LVT 
back to the current time.

The amount of time into the future that the emulation will attempt
to venture is another parameter which must be determined. This 
lookahead sliding window width, $\Lambda$, should be preconfigured 
based on the accuracy required; the farther ahead this predictive 
network management system attempts to predict 
past real time, the more risk that is assumed.

\subsection{Tolerance and Accumulated Simulation Error}

In order to consider the impact that out-of-tolerance rollback will 
have on the predictive system, consider how simulation error occurs.
A predictive management system Logical Process may deviate from the real object 
because either the Logical Process does not accurately represent the actual entity 
or because events outside the scope of the predictive network management 
system may effect the entities being managed. Ignore events outside the
scope of the simulation for now and consider error from inaccurate 
simulation modeling only. 

Because of the possibility for prediction error, a method of determining
the amount of error in a predicted result needs to be developed.
A function of total accumulated error in a predicted result, $AC(\cdot)$,
is described by the following Equations:

\begin{equation}
AC_n(n) = \sum_{i = 1}^N CE_{lp_i}(ME_{lp_{i-1}}, t_{lp_i})
\label{acn}
\end{equation}

\begin{equation}
AC_t(\tau) = \liminf_{\sum t_{lp_i} \rightarrow \tau}
\sum_{i = 1}^N CE_{lp_i}(ME_{lp_{i-1}}, t_{lp_i})
\label{actau}
\end{equation}

$ME_{dp}$ is the error introduced by the virtual message injected into the
predictive system by the driving process. The error introduced by the output 
message produced by the computation of each Logical Process is represented by 
the computation error function, $CE(\cdot)$. The actual time 
taken by the $n^{th}$ Logical Process to calculate and output the next virtual 
message is $t_{lp_n}$. Note that the Logical Process topology may not 
necessarily be a feed-forward network as described by Equations \ref{acn} and 
\ref{actau}; it may include a cycle. Note also that the right side of
Equation \ref{actau}
is the greatest lower bound of all sub-sequential limits of 
$\sum_{i = 1}^N CE_{lp_i}(ME_{lp_{i-1}}, t_{lp_i})$ as $\sum t_{lp_i} 
\rightarrow \tau$.

The driving process is indicated by $lp_0$.
$AC_n(n)$ is the total accumulated error in the virtual message output by
the $n^{th}$ Logical Process from the driving process. $AC_t(\tau)$ is the accumulated
error in $\tau$ actual time units from generation of the virtual message 
from the driving process. For example, if a prediction result is generated
in the third Logical Process from the driving process, then the total accumulated
error in the result is $AC_n(3)$. If 10 represents the number of time units
after the initial message was generated from the driving process then 
$AC_t(10)$ would be the amount of total accumulated error in the result.

\subsection{Optimum Choice of Verification Query Times}

As previously stated, the prototype system performs the 
verification based on the states in the state queue.
One method of choosing the verification query time is to query 
the managed entity based on the frequency of 
the data we are trying to monitor. Assuming the simulated data is
correct, query or sample in such a way as to perfectly reconstruct
the data, e.g. based on the maximum frequency component of the monitored
data. A possible
drawback is that the actual data may be changing at a multiple
of the predicted rate. The samples may appear to be accurate when
they are invalid.

\subsection{Verification Tolerance}

The verification tolerance, $\Theta$, is the amount of difference 
allowed between the Logical Process state and the actual entity state.
A large tolerance decreases the number of false messages and rollbacks,
thus increasing performance and requires fewer management queries. 
Allowing a larger probability of error between predicted and the actual
values will cause rollbacks in each Logical Process at real times of 
$t_{vfail}$ from the start of execution of each Logical Process.

The error throughout the simulated system may be randomized in such a way 
that errors among Logical Processes cancel. However, if the simulation is 
composed of many of the same class of Logical Process, the errors may compound 
rather than cancel each other. The tolerance of a particular Logical Process, 
$\Theta_{lp_n}$, will be reached in time $t_{vfail_n} = \{ \mbox{lub\ } \tau 
\mbox{ s.t. } AC_t(\tau) > \Theta_{lp_n} \}$.
The verification query period ($\Upsilon$) should be periodic with period 
less than or equal to $t_{vfail_n}$ in order to maintain accuracy within 
the tolerance.

The accuracy of any predicted event must be quantified. This could be
quantified as the probability of occurrence of a predicted event. The 
probability
of occurrence will be a function of the verification tolerance, the time
of last rollback due to verification error, the error between the 
simulation and actual entity, and the sliding lookahead window.
Every Logical Process will be in exact alignment with its Physical Process as a result of a 
state verification query. This occurs every
$T_{query} = t_{vfail}$ time units. 

\subsection{Length of Lookahead Window}

The length of the lookahead window, $\Lambda$, should be as large 
as possible while maintaining the required accuracy. The total error is 
a function of the chain of messages which lead to the state in 
question.
Thus, the farther ahead of real-time the predictive network management 
system advances,
$t_{ahead} = GVT - t_{current-time}$, the greater the number of 
messages before a verification query can be made and the 
greater the error. The maximum error is $AC_t(\Lambda)$.

\subsection{Simulation Time}

Since the verification query time is less than or equal
to the current time, $t_{current-time}$,
rollbacks due to the verification query will take the Logical Process 
back to the current time. Thus, Global Virtual Time as defined in \cite{Jefferson82}
is no longer a lower bound on the simulation rollback time. The lower
bound is now $t_{current-time}$. Global Virtual Time is still required in
order to determine how far into the future the predictive network 
management system has gone.

\subsection{Calibration Mode of Operation}

It may be helpful to run the predictive network management system in a mode 
such that error between the actual entities and the predictive network management system are 
measured. This error information can be used during the normal predictive
mode in order to help set the above parameters. This has an effect similar to 
back-propagation in a neural network, i.e. the predictive network 
management system automatically adjusts parameters in response to 
output in order to become more accurate. 
This calibration mode could be part of normal operation. The error can be
tracked simply by keeping track of the difference between the 
simulated messages and the result of verification queries. 
\section{MODEL AND SIMULATION}

The algorithm described in this paper has been implemented
and analyzed in \cite{BushThesis} that describes a predictive mobile
network. This paper extends the algorithm to network management.
An initial test of this algorithm in a network management environment
has been performed in a 
simulation\index{Sequential Simulation} of a predictive management system
implemented with Maisie \cite{bagrodia}. 
Its suitability has been demonstrated in the RDRN\index{RDRN}
network management and control design and development and in
\cite{Short} to develop a mobile wireless network
parallel simulation\index{Parallel Simulation} environment. The 
parallel simulation\index{Parallel Simulation} 
environment shows a speedup over the currently used commercial 
sequential simulation packages.  The environment 
and a set of modules which have been developed for mobile network
simulation are described in \cite{Short}.
Maisie uses a language which has been influenced by a classic work describing 
the characteristics of a parallel programming\index{Parallel Programming}
language structure \cite{Hoare81}. 

Since every Maisie 
entity has a built-in input queue, each Logical Process is comprised of three
additional Maisie entities:
\begin{itemize}
\item An entity which represents the Physical Process\index{Physical Process}
\item An entity for the Logical Process state queue\index{State Queue}
\item An entity for the Logical Process output message queue\index{Send Queue}
\end{itemize}

There is also a gvt entity for the calculation of Global Virtual Time\index{VNC!Global 
Virtual Time}. All three of the above entities work together to implement 
Virtual Time as described in \cite{Jefferson82}.
The first entity above, representing the Physical Process, 
contains a delay mechanism in order to implement the sliding 
lookahead window\index{Sliding Lookahead Window|see{LA}}.
The gvt process should notify all processes to cease forward simulation 
when Global Virtual Time reaches the end of the window. However, in this version of the predictive
management system, each Logical Process
simply compares its Local Virtual Time to the current time and holds processing
until current time is back within the lookahead sliding window.

Determination of Global Virtual Time should be done as defined by
\cite{Lazowaska90}.
This algorithm allows Global Virtual Time to be determined in a message-passing environment
as opposed to the easier case of a shared memory environment. It also
allows normal processing to continue during the Global Virtual Time determination phase.
However, in this implementation each output message is sent to the gvt 
entity as well as to its proper destination. In addition, the gvt entity 
checks all Logical Processes for their current Local Virtual Time and chooses the minimum message send time
and Local Virtual Time as the current Global Virtual Time. The gvt entity is allowed to execute in 
parallel with other entities in this simulation, thus it may not always be 
perfectly accurate. This is because messages may be in transit when
the Local Virtual Time request poll takes place, and because Logical Processes are changing 
while the Global Virtual Time computation is taking place.
However, the results are close enough for the purpose
of these experiments.

\subsection{Verification Query Rollback Versus Causality Rollback}

Verification query\index{State Adjustment|see{VNC}} rollbacks are the 
most critical part of the predictive management system. They are handled in a slightly different fashion 
from causality\index{VNC!causality} failure rollbacks. A state verification 
failure causes the Logical Process state to be corrected at the time of the state 
verification that failed. 
The state, $S_{v}$, has been obtained from the actual device from 
the verification query at time $t_{v}$.
The Logical Process rolls back to exactly $t_{v}$ with state, $S_{v}$.
States greater than $t_{v}$ are removed from the state 
queue\index{State Queue}. 
Anti-messages\index{Anti-Message} are sent from the output message queue 
for all messages
greater than $t_{v}$. The Logical Process continues forward execution from this 
point.
Note that this implies that the message and state queues cannot
be purged of elements that are older than the Global Virtual Time. Only elements which
are older than real time can be purged.

\subsection{The Prototype System Simulation}

\begin{figure*}[htbp]
\centerline{\psfig{file=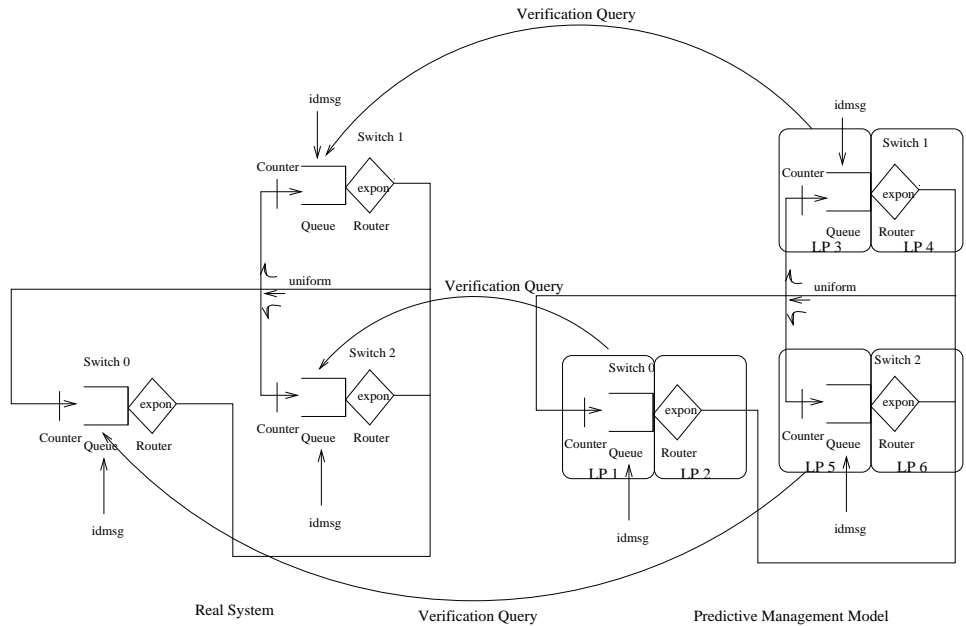,width=5.0in}}
\caption{Initial Feasibility Network Model}
\label{qrsim}
\end{figure*}

A simple network management system was simulated to test the concept
of the predictive management protocol just described. Note that none of the
previous assumptions are made in the simulation.
The purpose of this simulation is to determine if the concept
is feasible\index{VNC!feasibility simulation}. A key question this simulation 
attempts to answer is whether the overhead/performance ratio results in a 
useful system. A small closed queuing network with First Come First Serve servers
represents the actual system. Figure \ref{qrsim} shows the
real system to be managed and the predictive management model. In this
initial feasibility study, the managed system and the predictive management
model are both modeled with Maisie. The verification query between the
real system and the management model are explicitly illustrated in
Figure \ref{qrsim}.

The system consists of three switch-like entities, each switch contains a 
single queue and switches consisting of 10 exponentially distributed servers 
that sequentially service each packet. A mean service time of 10 time 
units is
assumed. The servers represent the link rate. The packet is then forwarded 
with equal probability to another switch, including the originating switch.
Each switch is a driving process; the switches forward real and virtual 
messages.
The cumulative number of packets which have entered each switch's queue 
is the state. This is similar to Simple Network Management Protocol \cite{SNMP} statistics monitored
by Simple Network Management Protocol Counters, for example, the {\bf ifInOctets} counter in
MIB-II interfaces \cite{RFC1156}.

Both real and virtual messages contain the time service ends and 
a count of the number of times a packet has entered a switch. 
An initial message 
enters each queue upon startup to associate a queue with its switch.
This is the purpose of the {\bf idmsg} that enters the queues in
Figure \ref{qrsim}. The predictive system parameters 
are more compactly identified as a triple consisting of
Lookahead Window Size (seconds),  Tolerance (counter value), 
and Verification Query Period (seconds) in the form $(\Lambda, 
\Theta, \Upsilon)$. The effect of these parameters are examined on
the system of switches previously described.
The simulation was run with the following triples:
$(5,10,5)$, $(5,10,1)$, $(5,3,5)$, $(400,5,5)$. The graphs that follow
show the results for each triple. 

\begin{figure*}[htbp]
\centerline{\psfig{file=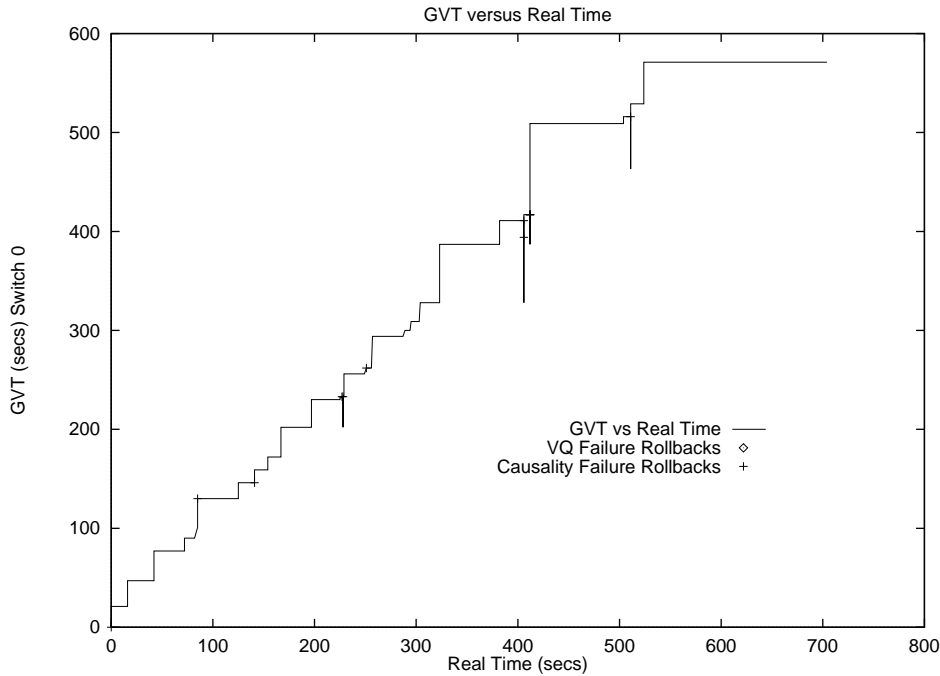,width=5.0in}}
\caption{Rollbacks Due to State Verification Failure (5, 10, 5)}
\label{rb5105}
\end{figure*}

\begin{figure*}[htbp]
\centerline{\psfig{file=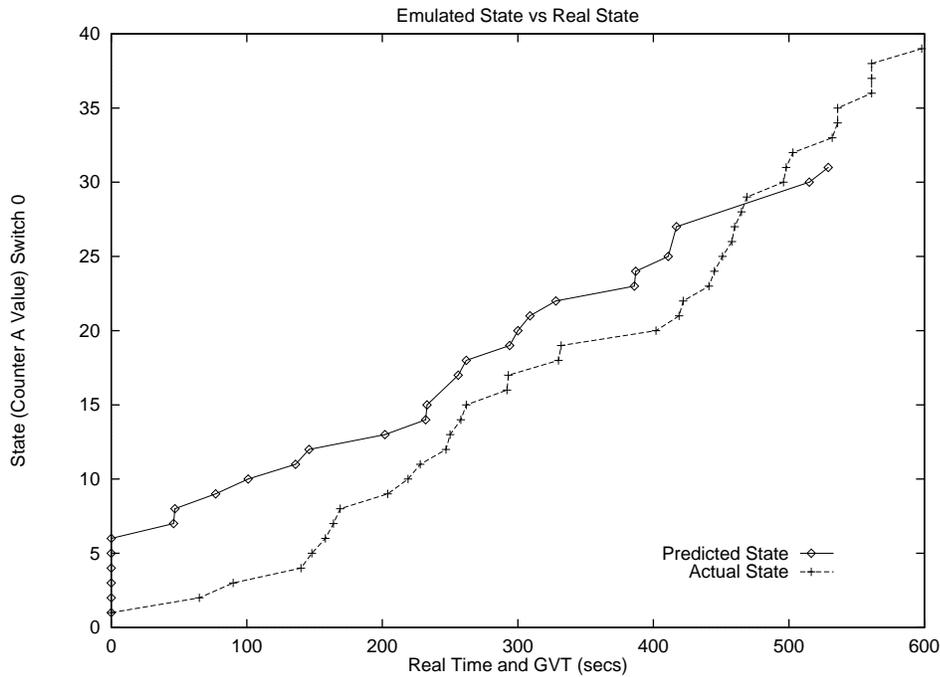,width=5.0in}}
\caption{State (5, 10, 5)}
\label{s5105}
\end{figure*}

The first run parameters were $(5, 10, 5)$. There were no state 
verification rollbacks although there were some causality induced rollbacks
as shown in Figure \ref{rb5105}. Global Virtual Time increased almost instantaneously versus 
real time; at times the next event far exceeded the look-ahead window. This
is the reason for the nearly vertical jumps in the Global Virtual Time as a function of 
real-time as shown in Figure \ref{rb5105}. The state graph for this
run is shown in Figure \ref{s5105}.

In the initial implementation, state verification was performed in the
Logical Process immediately after each new message was received. However, the probability
that an Logical Process had saved a future state, while processing at its Local Virtual Time,
with the same state save time as the time at which a real message arrived was low. 
Thus, there was frequently nothing to 
compare the current state with in order to perform the state verification.
However, it was observed that the predictive system was simulating up to the lookahead 
window very quickly and spending most of its time holding, during which 
time it was doing nothing. The implementation was modified so that each entity 
would perform state verification during its hold time\index{VNC!hold time}.
This design change better utilized the processors and resulted in
more accurate alignment between actual and logical processes.

\begin{figure*}[htbp]
\centerline{\psfig{file=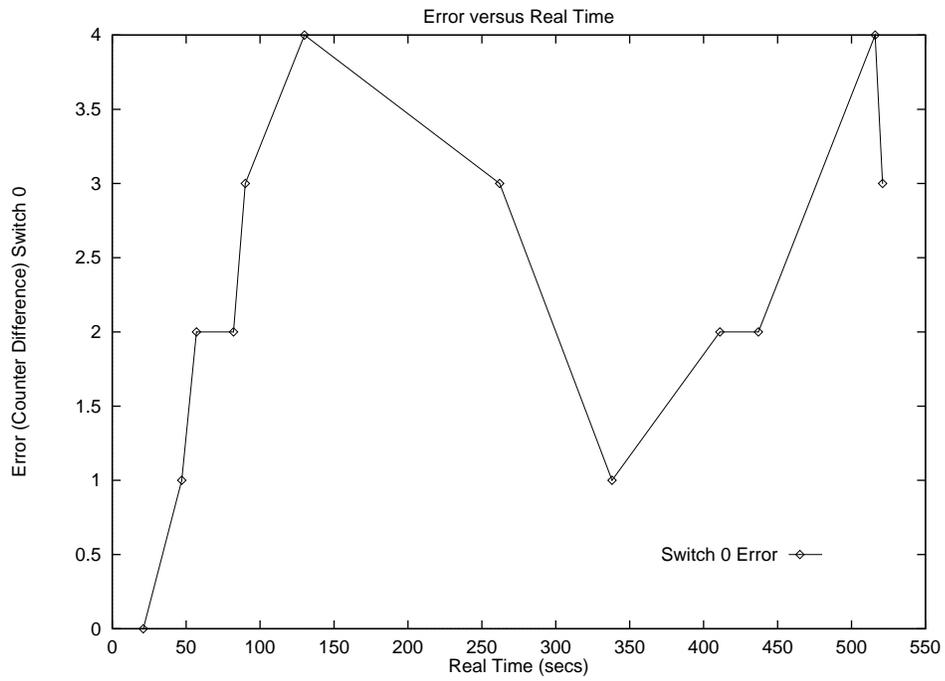,width=5.0in}}
\caption{Amount of Error (5, 10, 5)}
\label{e5105}
\end{figure*}

\begin{figure*}[htbp]
\centerline{\psfig{file=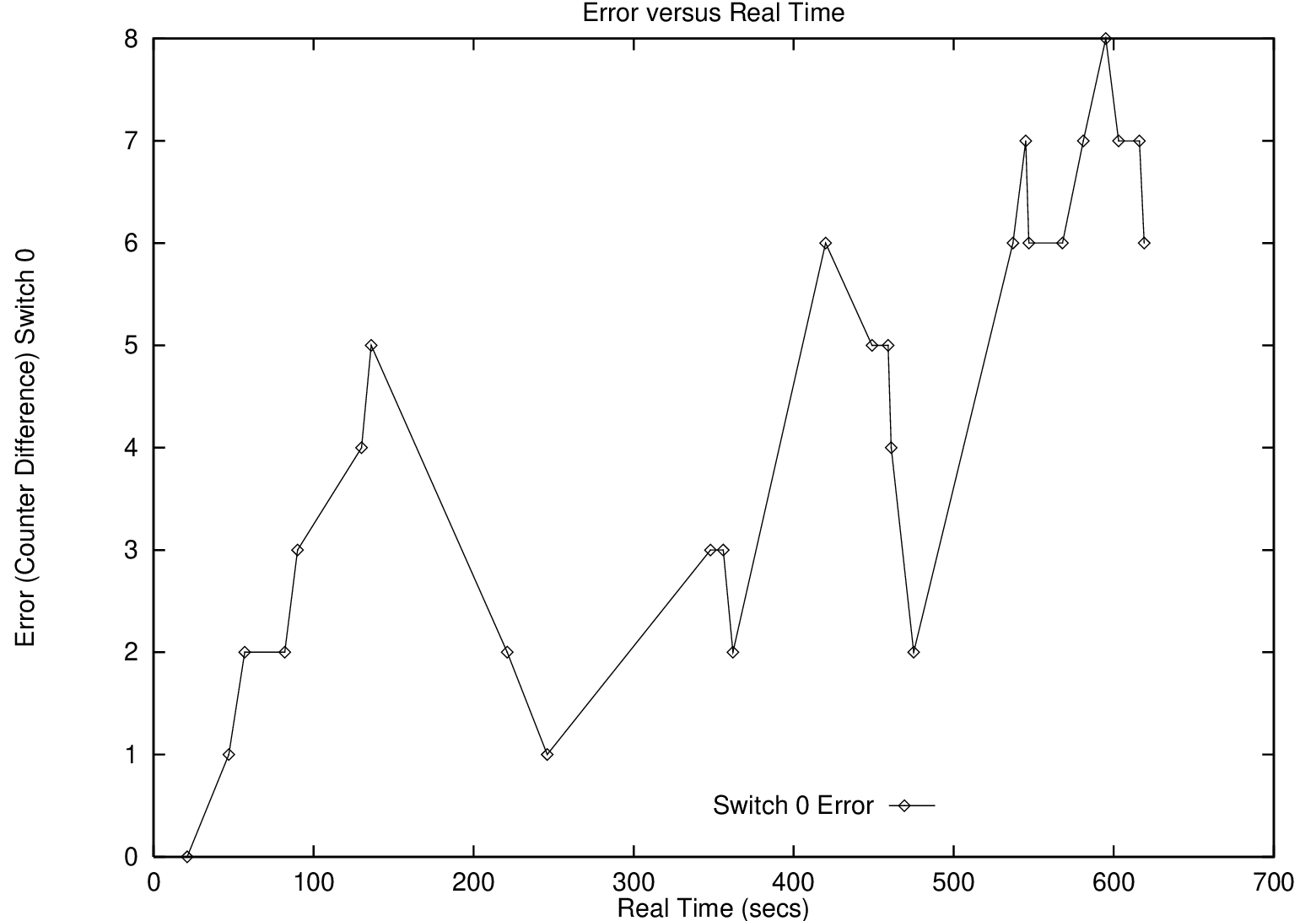,width=5.0in}}
\caption{Amount of Error (5, 10, 1)}
\label{e5101}
\end{figure*}

\begin{figure*}[htbp]
\centerline{\psfig{file=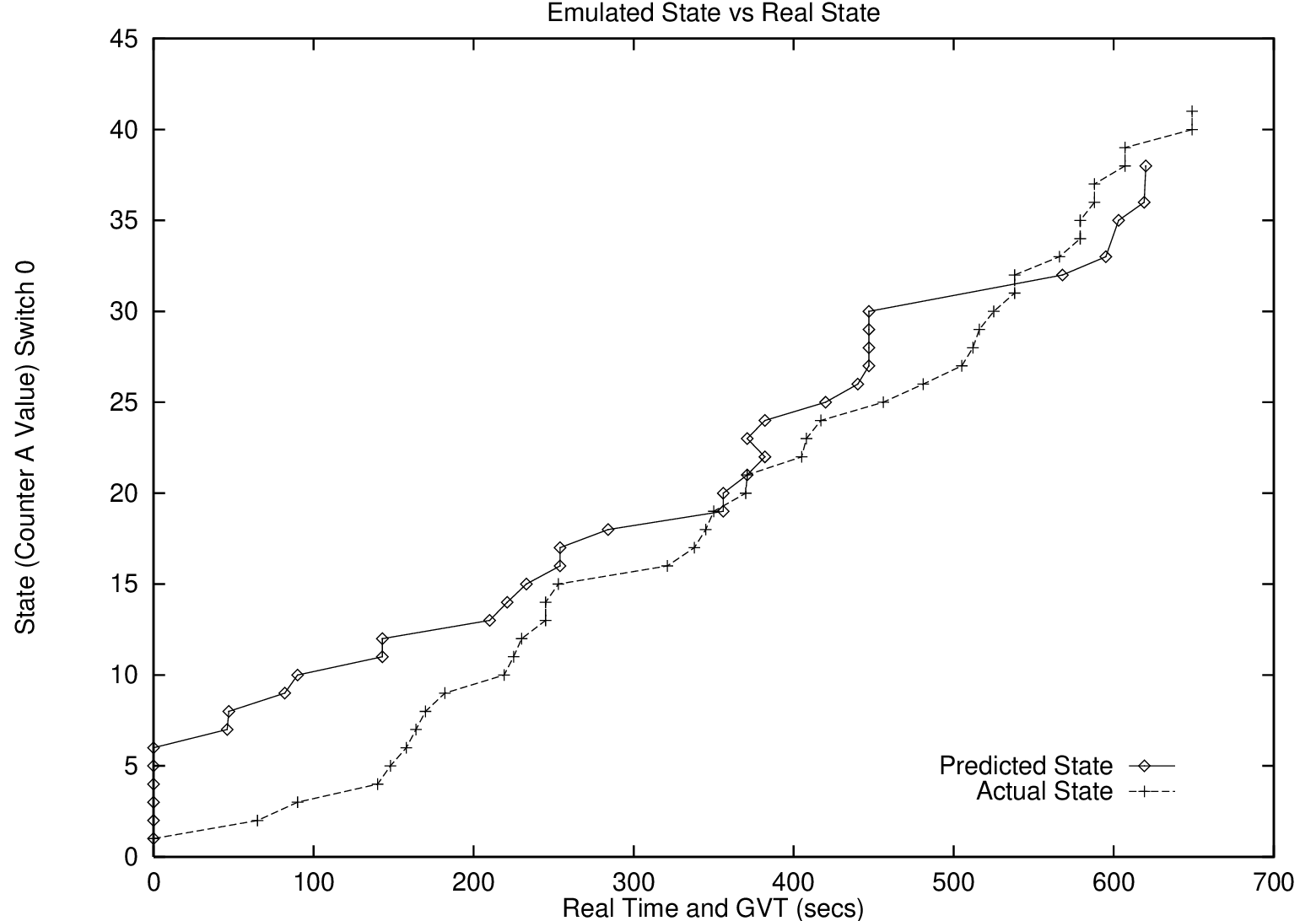,width=5.0in}}
\caption{State (5, 10, 1)}
\label{s5101}
\end{figure*}

The results for the $(5, 10, 1)$ run were similar, except that the 
predictive and actual system comparisons were more frequent because the state 
verification period had been changed from once every 5 seconds to once 
every second. Error was measured as the difference in the predicted Logical Process state 
versus the actual system state. 
This run showed errors that were greater than those in the first run, great 
enough to cause state verification rollbacks.
The error levels for both runs are shown in
Figures \ref{e5105} and \ref{e5101}. The state graph for this run is
shown in Figure \ref{s5101}.

\begin{figure*}[htbp]
\centerline{\psfig{file=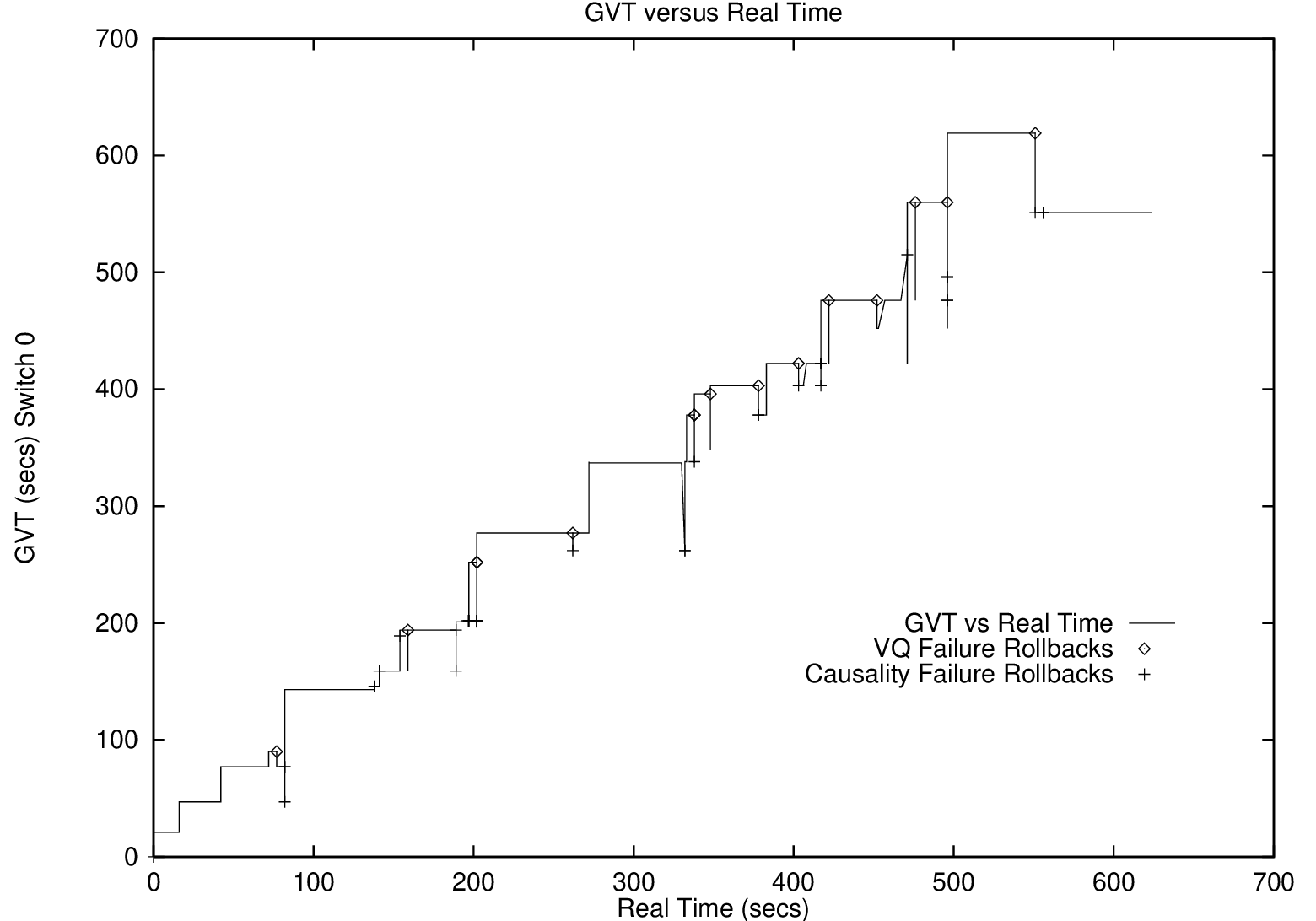,width=5.0in}}
\caption{Rollbacks Due to State Verification Failure (5, 3, 5)}
\label{rb535}
\end{figure*}

\begin{figure*}[htbp]
\centerline{\psfig{file=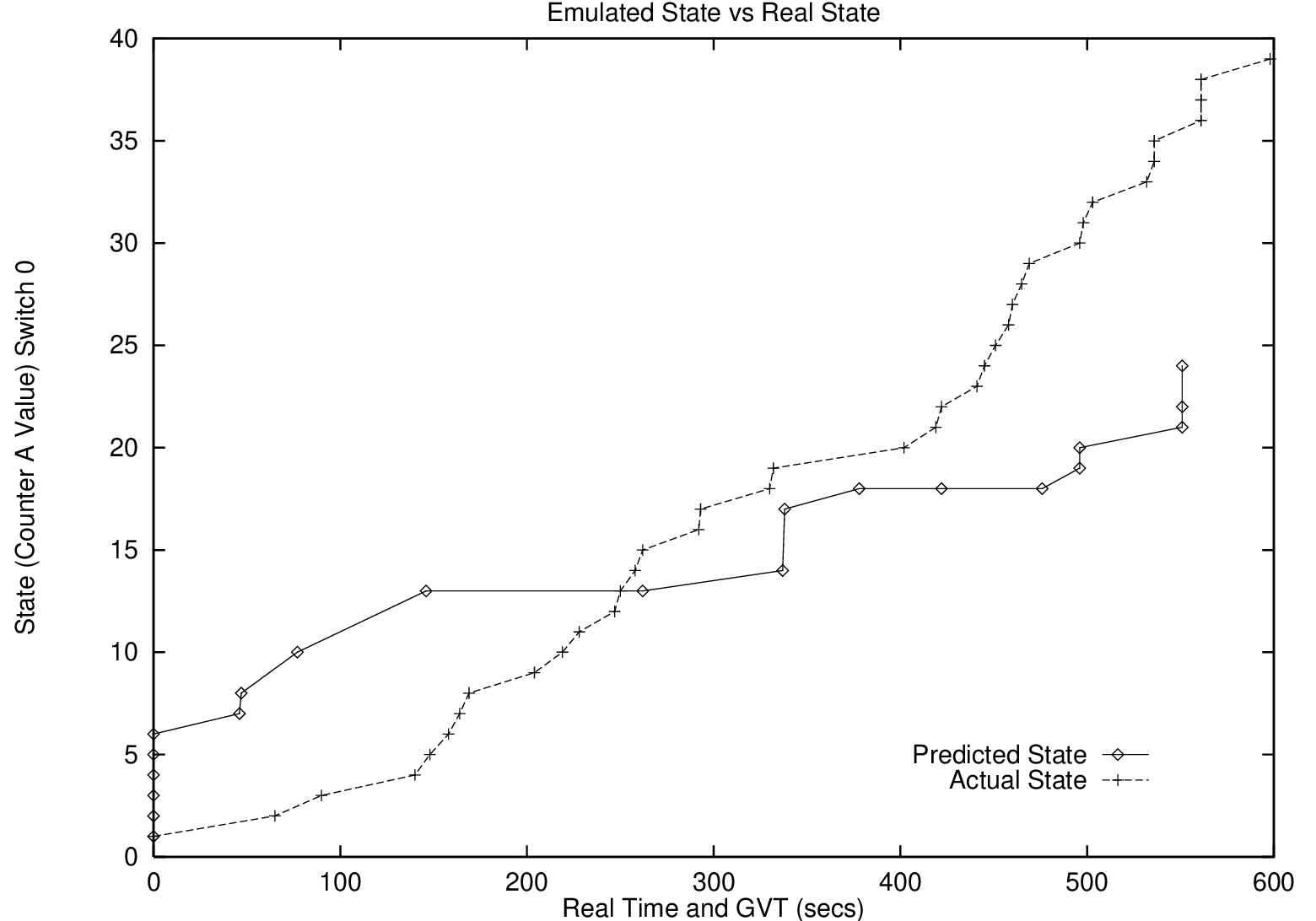,width=5.0in}}
\caption{State (5, 3, 5)}
\label{s535}
\end{figure*}

The next run used $(5, 3, 5)$ parameters. Here we see many more state 
verification failure rollbacks as shown in Figure \ref{rb535}. This is
expected since the tolerance has been reduced from 10 to 3. The cluster of  
causality rollbacks near the state verification rollbacks was expected.
These clusters of causality rollbacks do not appear to significantly           
reduce the feasibility of the system. The real-time 
versus Global Virtual Time plot as shown in Figure \ref{rb535} shows much 
larger jumps as the Logical Processes were held back due to rollbacks. The entities 
had a larger variance in their hold times than the $(5, 10, 5)$ run. The
state graph for this run is shown in Figure \ref{s535}. 

\begin{figure*}[htbp]
\centerline{\psfig{file=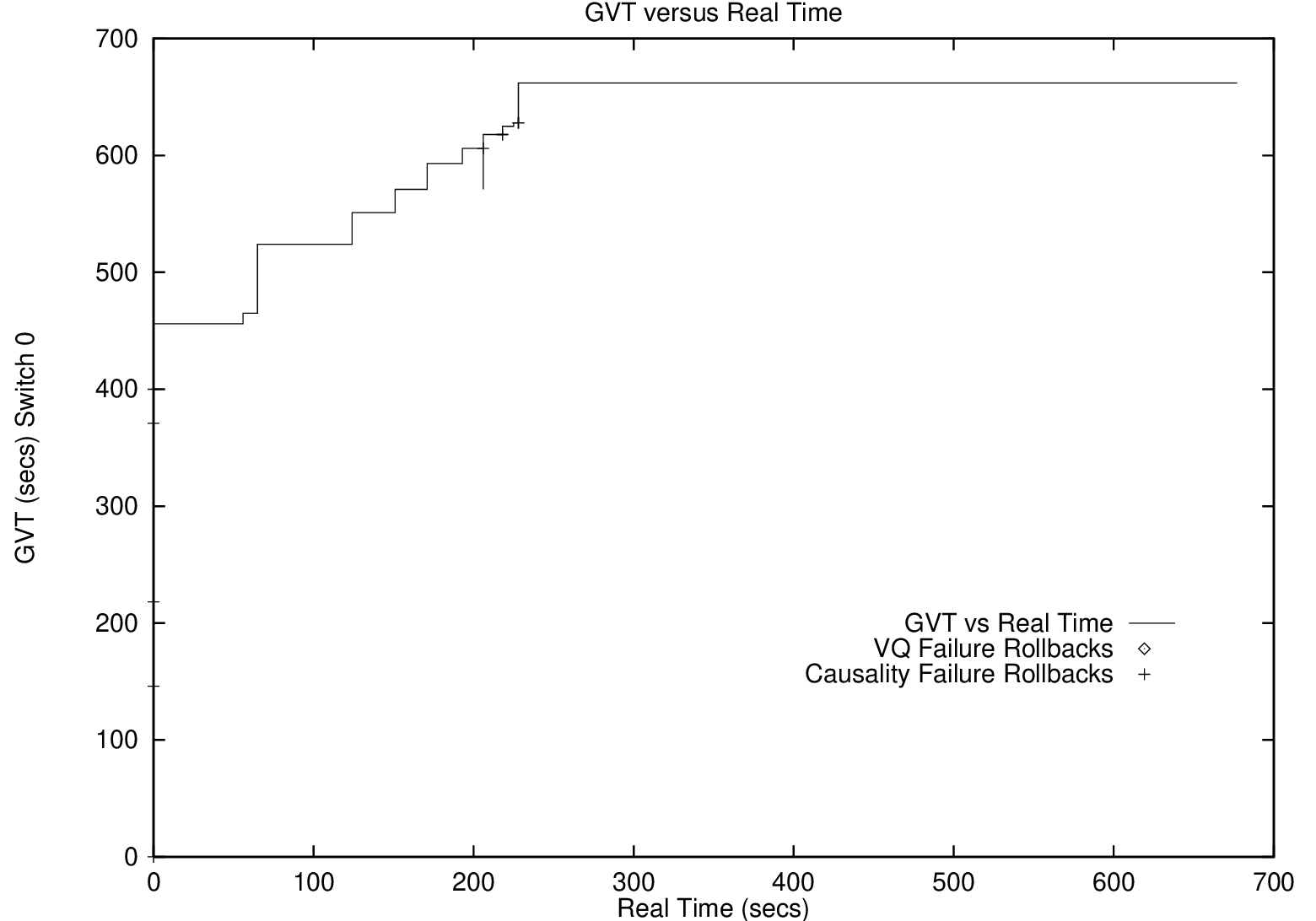,width=5.0in}}
\caption{Rollbacks Due to State Verification Failure (400, 5, 5)}
\label{rb40055}
\end{figure*}

\begin{figure*}[htbp]
\centerline{\psfig{file=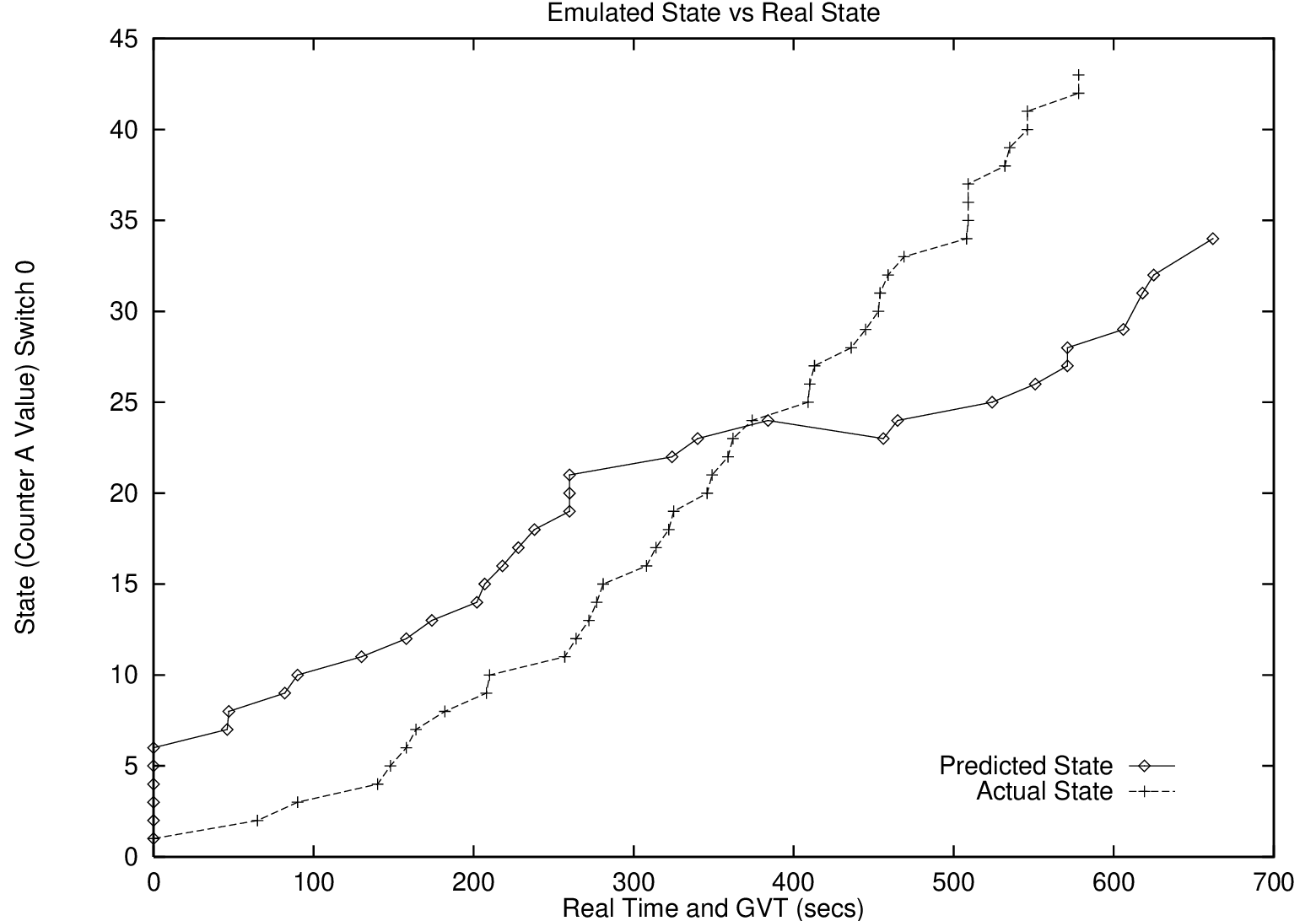,width=5.0in}}
\caption{State (400, 5, 5)}
\label{s40055}
\end{figure*}

A $(400, 5, 5)$ run showed the Global Virtual Time jump quickly to 400 and then gradually
increase as the sliding lookahead window maintained a 400 time unit lead
as shown in Figure \ref{rb40055}.
The Logical Process hold times were shorter than an any previous run. The state
graph for this run is shown in Figure \ref{s40055}. 

This set of results is interesting because it shows the system to be
stable with the introduction of state verification rollbacks. The
overhead introduced by these rollbacks did not greatly impact the
performance, because as previously shown in the Global Virtual Time versus time graphs
in Figures \ref{rb5105}, \ref{rb535}, and \ref{rb40055}, the system was 
always able to predict up to its lookahead time very quickly.
\section{OPTIMIZING MANAGEMENT POLLING WITH THE PREDICTIVE MANAGER}

Since the predictive network management system provides a good 
approximation of the future behavior of the
data to be managed as shown in the Global Virtual Time versus real time values of state 
in Figures \ref{s5105}, \ref{s5101}, \ref{s535}, and \ref{s40055}, the 
verification query period can be automatically
determined as a function of the look-ahead window and tolerance, with
the goal of minimizing the frequency of verification queries thus
solving the polling problem in network management.

In most standards based approaches, network management stations
are sampling counters in managed entities that 
simply increment in value until they roll over. A management station
which is simply plotting data will have some fixed polling interval
and record the absolute value of the difference in value of the
counter. Such a graph is not a perfectly accurate representation
of the data, it is merely a statement that sometime within a polling
interval the counter has monotonically increased by some amount.
Spikes in this data, which may be very important to the current state of the
system, may not be noticed if the polling interval
is long enough such that a spike followed by low data values averages out
to a normal or low value. One of the goals of a predictive management
system is to determine the minimum
polling interval required to accurately represent the data.

From the information provided by the predictive
management system, a polling interval which provides the desired
degree of accuracy can be determined and dynamically adjusted; however, 
the cost must be determined. 
An upper limit on the number of systems that can be polled
is $N \le {T \over \Delta}$ where $N$ is the number of devices capable of
being polled, $T$ is the polling interval, and $\Delta$ is the time
required for a single poll. Thus although the data accuracy will be
constrained by this upper limit, taking advantage of characteristics
of the data to be monitored can help distribute the polling intervals
efficiently within this constraint.
Assume that $\Delta$ is a calculated and fixed value, as is $N$.
Thus this is a lower bound on the value of $T \ge \Delta N$.

The overhead\index{Virtual Network Configuration!overhead} bandwidth
required for use by the management system to
perform polling is shown in Equation \ref{bw}. The packet size will
vary depending upon whether it is an SNMP or CMIP packet
and the MIB
object(s) being polled. The number of packets varies with the amount
of management data requested. Let $P$ be the number of packets,
$S$ be the bits/packet, $N$ be the number of devices polled, and $T$
be the polling period. $Bw$ is the total available bandwidth and
$Bw_{oh}$ is the overhead\index{Virtual Network Configuration!overhead}
bandwidth of the management traffic.

\begin{figure*}
\begin{equation}
Bw_{oh}\% = {100.0 P N S Bw \over T}
\label{bw}
\end{equation}
\end{figure*}

It may be desirable to limit the bandwidth used for polling system
management
data to be no more than a certain percentage of total bandwidth. Thus
the optimum polling interval will use the least amount of bandwidth
while also maintaining the least amount of variance due to error in the
data signal. All the required information to maintain the cost versus
accuracy at a desired level is provided by the
predictive\index{Predictive Network Management!predictive} network
management system.
\section{INTERACTION BETWEEN A PREDICTIVE MANAGEMENT SYSTEM AND A
PREDICTIVE MOBILE NETWORK}

There is an interesting interaction between the predictive management
system and the predictive mobile network. A predictive mobile network
such as the Rapidly Deployable Radio Network \cite{BushICC96, BushThesis} 
will have cached results
in advance of use for many configuration parameters. 
These results should be part of the Management Information Base 
for the mobile network and includes the predicted time of the
event requiring the result, the value of the result, and the
probability that the result will be within tolerance. Thus there
will be a triple associated with each predicted event: 
{\em (time, value, probability)}. Network management protocols, e.g. SNMP
\cite{SNMP} and CMIP \cite{CMIP}, include the time as part of the Protocol Data Unit, 
however this time indicates the real time the poll occurred.

A predictive management system could simply use Logical Processes to represent the
predictive mobile processes as previously described, however, this is
redundant since the mobile network itself has predicted events in
advance as part of its own management and control system. Therefore,
managing a predictive mobile network with a predictive network
management system provides an interesting problem in trying
to get the maximum benefit from both of these predictive systems.

Combining the two predictive systems in a low level manner, e.g. allowing 
the Logical Processes to exchange messages with each other, raises questions about 
synchronization between the mobile network and the management station. 
However, the predicted
mobile network results can be used as additional information to
refine the management system results. The management system will have
computed {\em (time, value, probability)} triples for each predicted result
as well. The final result by the management system would then be
an average of the {\em times} and {\em values} weighted by their 
respective probabilities. 
An additional weight may be added given the quality of either system. For
example the network management system might be weighted higher because
it has more knowledge about the entire network. Alternatively, the mobile 
network system may weighted higher because the mobile
system may have better predictive capability for the detailed
events concerning handoff. Thus the two systems do not directly interact
with each other, but the final result is a combination of the results from
both predictive systems. A more complex method of combining results from these
two systems would involve a causal network such as the one described in 
\cite{Lehmann}.
\section{CONCLUSION}

Network management systems capable not only of passive monitoring
but also of active prediction capability are undergoing
research and development. Work on prediction
mechanisms for mobile communication networks is also underway.
A method used by standards-based network management systems
to cope with these two developments has been proposed in this paper.

A predictive network management algorithm 
and the characteristics of a predictive network management 
system have been presented. The predictive capability of the
network management system is used to solve the 
polling rate problem for network management. The Rapidly Deployable
Radio Network \cite{BushThesis} is presented as an example of a predictive
mobile communications network. Finally, interaction between the predictive
capabilities of the network management and mobile network systems 
has been discussed.
%
%
\medskip
\bibliographystyle{unsrt} 
\bibliography{/home/bushsf/ref/standards,/home/bushsf/ref/topo,/home/bushsf/ref/psim,/home/bushsf/ref/mob,/home/bushsf/ref/twe}
\end{document}